\DeclareSymbolFont{matha}{OML}{txmi}{m}{it}
\DeclareMathSymbol{\varv}{\mathord}{matha}{118}

\documentclass[twoside,twocolumn,9pt]{article}
\usepackage[super,sort&compress,comma]{natbib} 
\usepackage[version=3]{mhchem}
\usepackage[left=1.5cm, right=1.5cm, top=1.785cm, bottom=2.0cm]{geometry}
\usepackage{balance}
\usepackage{mathptmx}
\usepackage{sectsty} 
\usepackage{graphicx} 
\usepackage{lastpage}
\usepackage[format=plain,justification=justified,singlelinecheck=false,font={stretch=1.125,small,sf},labelfont=bf,labelsep=space]{caption}
\usepackage{float}
\usepackage{fancyhdr}
\usepackage{fnpos}
\usepackage[english]{babel}
\addto{\captionsenglish}{%
  \renewcommand{\refname}{Notes and references}
}
\usepackage{array}
\usepackage{droidsans}
\usepackage{charter}
\usepackage[T1]{fontenc}
\usepackage[usenames,dvipsnames]{xcolor}
\usepackage{setspace}
\usepackage[compact]{titlesec}
\usepackage{hyperref}
\definecolor{magenta}{rgb}{1.0, 0.0, 1.0} 
\definecolor{green}{rgb}{0.0, 1.0, 0.0} 
\definecolor{dodgerblue}{rgb}{0.12, 0.56, 1.0} 
\definecolor{navy}{rgb}{0.0, 0.0, 0.5} 
\definecolor{coral}{rgb}{1.0, 0.5, 0.31} 
\definecolor{cyan}{rgb}{0.0, 1.0, 1.0} 
\definecolor{pink}{rgb}{1.0, 0.75, 0.8} 
\definecolor{orange}{rgb}{1.0, 0.647, 0.0} 
\definecolor{purple}{rgb}{0.5, 0.0, 0.5} 
\definecolor{blue}{rgb}{0.0, 0.0, 1.0} 
\definecolor{red}{rgb}{1.0, 0.0, 0.0} 
\definecolor{lime}{rgb}{0.0, 1.0, 0.0} 
\definecolor{darkgreen}{rgb}{0.0, 0.39, 0.0} 
\usepackage{xcolor}     
\usepackage{amsmath}    
\usepackage{amssymb}    

\usepackage{epstopdf}

\definecolor{cream}{RGB}{222,217,201}

\usepackage{placeins}
\usepackage[normalem]{ulem}

    \newcommand{\blu}[1]{{\color{black}{#1}}}
    \newcommand{\panos}[1]{{\color{black}{#1}}}

    \newcommand{\vio}[1]{{\color{black}{#1}}}
    \newcommand{\vioo}[1]{{\color{black}{#1}}}
    \newcommand{\viooo}[1]{{\color{black}{#1}}}
    \newcommand{\pd}[1]{{\begin{color}[rgb]{0,0,0}{#1}\end{color}}}

\newcommand*{\citen}[1]{
  \begingroup
    \romannumeral-`\x 
    \setcitestyle{numbers}%
    \cite{#1}%
  \endgroup   
}

\begin{document}

\pagestyle{fancy}
\thispagestyle{plain}
\fancypagestyle{plain}{
\renewcommand{\headrulewidth}{0pt}
}

\makeFNbottom
\makeatletter
\renewcommand\LARGE{\@setfontsize\LARGE{15pt}{17}}
\renewcommand\Large{\@setfontsize\Large{12pt}{14}}
\renewcommand\large{\@setfontsize\large{10pt}{12}}
\renewcommand\footnotesize{\@setfontsize\footnotesize{7pt}{10}}
\makeatother

\renewcommand{\thefootnote}{\fnsymbol{footnote}}
\renewcommand\footnoterule{\vspace*{1pt}%
\color{cream}\hrule width 3.5in height 0.4pt \color{black}\vspace*{5pt}} 
\setcounter{secnumdepth}{5}

\makeatletter 
\renewcommand\@biblabel[1]{#1}            
\renewcommand\@makefntext[1]%
{\noindent\makebox[0pt][r]{\@thefnmark\,}#1}
\makeatother 
\renewcommand{\figurename}{\small{Fig.}~}
\sectionfont{\sffamily\Large}
\subsectionfont{\normalsize}
\subsubsectionfont{\bf}
\setstretch{1.125} 
\setlength{\skip\footins}{0.8cm}
\setlength{\footnotesep}{0.25cm}
\setlength{\jot}{10pt}
\titlespacing*{\section}{0pt}{4pt}{4pt}
\titlespacing*{\subsection}{0pt}{15pt}{1pt}

\fancyfoot{}
\fancyfoot[LO,RE]{\vspace{-7.1pt}\includegraphics[height=9pt]{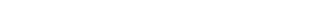}}
\fancyfoot[CO]{\vspace{-7.1pt}\hspace{13.2cm}\includegraphics{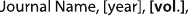}}
\fancyfoot[CE]{\vspace{-7.2pt}\hspace{-14.2cm}\includegraphics{head_foot/RF}}
\fancyfoot[RO]{\footnotesize{\sffamily{1--\pageref{LastPage} ~\textbar  \hspace{2pt}\thepage}}}
\fancyfoot[LE]{\footnotesize{\sffamily{\thepage~\textbar\hspace{3.45cm} 1--\pageref{LastPage}}}}
\fancyhead{}
\renewcommand{\headrulewidth}{0pt} 
\renewcommand{\footrulewidth}{0pt}
\setlength{\arrayrulewidth}{1pt}
\setlength{\columnsep}{6.5mm}
\setlength\bibsep{1pt}

\makeatletter 
\newlength{\figrulesep} 
\setlength{\figrulesep}{0.5\textfloatsep} 

\newcommand{\topfigrule}{\vspace*{-1pt}%
\noindent{\color{cream}\rule[-\figrulesep]{\columnwidth}{1.5pt}} }

\newcommand{\botfigrule}{\vspace*{-2pt}%
\noindent{\color{cream}\rule[\figrulesep]{\columnwidth}{1.5pt}} }

\newcommand{\dblfigrule}{\vspace*{-1pt}%
\noindent{\color{cream}\rule[-\figrulesep]{\textwidth}{1.5pt}} }

\makeatother

\twocolumn[
  \begin{@twocolumnfalse}
{\includegraphics[height=30pt]{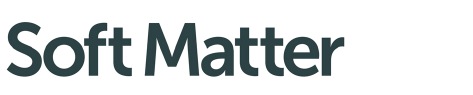}\hfill\raisebox{0pt}[0pt][0pt]{\includegraphics[height=55pt]{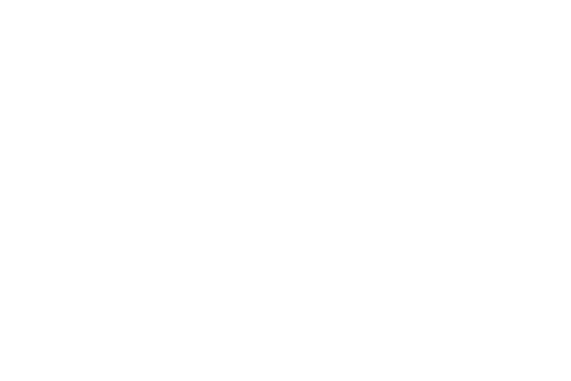}}\\[1ex]
\includegraphics[width=18.5cm]{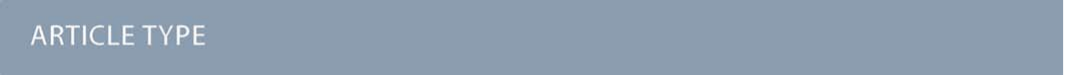}}\par
\vspace{1em}
\sffamily
\begin{tabular}{m{4.5cm} p{13.5cm} }

\includegraphics{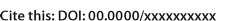} & \noindent\LARGE{\textbf{Collision of surfactant-laden droplets: Insights from molecular dynamics simulation$^\dag$}} \\
\vspace{0.3cm} & \vspace{0.3cm} \\

 & \noindent\large{Soheil Arbabi,\textit{$^{a\, \, b}$} Piotr Deuar,\textit{$^{a}$} Rachid Bennacer,\textit{$^{c}$}  Zhizhao Che,\textit{$^{d}$} and Panagiotis E. Theodorakis\textit{$^{a}$}} \\

\includegraphics{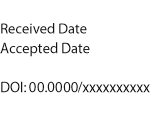} & \noindent\normalsize{We study the collision dynamics of surfactant-laden droplets and compare it with that of pure water droplets,
with a focus on bridge growth rate, energy balance, and disk dynamics, 
distinguishing the cases of head-on and off-centre collisions.
By using molecular dynamics simulation of a coarse-grained model, 
it is found that an initial linear scaling
describes the first stage of the collision 
process, which is followed by power-law dynamics, in 
contrast to an initial thermal regime and a subsequent
power-law behaviour observed for droplet coalescence. 
The transition between the two regimes 
occurs faster for surfactant-laden droplets. 
At higher collision velocities, the linear regime
dominates the process with a gradual reduction of the
power-law behaviour, reaching a situation in which
the bridge growth is fully characterised by linear 
dynamics. The different behaviour of the droplets
is presented in the form of a diagram of different scenarios,
namely coalescence, separation, and splattering. 
In particular, it is found that higher velocities
and larger offsets increase the likelihood of separation 
and splattering, with water droplets producing a 
greater number of satellite droplets due to reduced 
viscous damping. 
Also, a disk-like structure is observed as a result
of collision, but it is less pronounced in the case of surfactant-laden 
droplets, due to higher dissipation of energy. 
}

\end{tabular}

 \end{@twocolumnfalse} \vspace{0.6cm}

  ]

\renewcommand*\rmdefault{bch}\normalfont\upshape
\rmfamily
\section*{}
\vspace{-1cm}


\footnotetext{\textit{$^{a}$~Institute of Physics, Polish Academy of Sciences, Al. Lotnik\'ow 32/46, 02-668 Warsaw, Poland. E-mail: arbabi@ifpan.edu.pl
}}
\footnotetext{\textit{$^{b}$~Department of Biomedical
Engineering, University
of Wisconsin-Milwaukee,
Milwaukee, WI 53211, USA. 
}}
\footnotetext{\textit{$^{c}$~Université Paris-Saclay, CentraleSupélec, ENS Paris-Saclay, CNRS, LMPS - Laboratoire de Mécanique Paris-Saclay, 91190, Gif-sur-Yvette, France}}
\footnotetext{\textit{$^{d}$~State Key Laboratory of Engines, Tianjin University, 300350 Tianjin, China. }}



\section{Introduction}
\label{intro}
Droplet collision and coalescence are 
ubiquitous phenomena in nature, which, for example,
manifest in clouds and raindrops.\cite{falkovich2002acceleration} 
At the same time these are relevant for applications
in industry, such as liquid-fuel combustion 
and spray technology.\cite{Williams1990}
In view of the interest in these phenomena, 
various studies have been conducted towards obtaining a better
understanding of them, 
but most of these
investigations have thus far mainly
focused on pure water or polymer droplets.\cite{krishnan2015effects,yoon2007coalescence,khodabocus2018scaling,perumanath2019droplet,eggers1999coalescence,aarts2005hydrodynamics,sprittles2012coalescence,Dudek2020,Rahman2019,Berry2017,Somwanshi2018,Kirar2020,Bayani2018,Brik2021,Anthony2020,Kern2022,Heinen2022,Geri2017,Abouelsoud2021,Dekker2022,Calvo2019,Sivasankar2022,Otazo2019,Vannozzi2019,Arbabi2023Spolymer}
Although understanding collision and coalescence in
the context of pure liquids is fundamental, most
processes in industry (\textit{e.g.} enhanced oil recovery\cite{Massarweh2020}) as well as natural processes (\textit{e.g.} formation of clouds\cite{Denys2022}) take place in the presence
of surfactant. Coalescence of such droplets has
been previously explored by using molecular dynamics 
simulations of freely 
suspended\cite{arbabi2023SoftMatter,arbabi2023bcoalescencePhysFluid}
or sessile droplets,\cite{arbabi2024SessilePhysFluid}
since this method can provide a molecular-level resolution of
the relevant processes, such as
the surfactant mass transport mechanism. In this regard,
a detailed description of the mass transport mechanism 
and bridge growth dynamics has been provided.\cite{arbabi2023SoftMatter,arbabi2023bcoalescencePhysFluid,arbabi2024SessilePhysFluid}
However, in various applications and phenomena,
coalescence takes place while droplets move against each other
with a relative velocity,
in which case  
inertial effects are likely to
play a more
dominant role in the process, for example, by suppressing
the initial thermal 
regime. \cite{arbabi2023SoftMatter,arbabi2023bcoalescencePhysFluid}
Currently, there is a lack of insight into the coalescence of
colliding surfactant-laden droplets from a molecular perspective, 
which calls for further research in this area, especially
in view of the interest in this process from
a fundamental as well as an industrial perspective.


\panos{Several studies have overall provided valuable insights into the 
collision of droplets, especially for the case of
pure liquids. \pd{In particular}, 
Zhang \textit{et al.} \cite{zhang2019regimes} conducted molecular dynamics
simulations of head-on collisions of water droplets with a 
diameter of $10.9\,~\text{nm}$ for a wide range of Weber 
numbers, $We={\rho V^2 D_0}/{\gamma}$, where $\rho$ is the liquid density,
$V$ the velocity of droplets, $D_0$ the droplet diameter,
and $\gamma$ the surface tension. This study has suggested that when the ratio 
of the expansion disk diameter to the droplet's initial diameter
($D_d/D_0$) is around 2.66, the liquid film becomes unstable,
thus leading to rupture and eventually to the formation of cavities within the film.
In this way, the initial kinetic energy of droplets is 
dissipated through viscous dissipation during the collision. 
Moreover, larger viscosity would lead to greater energy 
dissipation, which could preserve the integrity of the expanding disk, 
that is to prevent the occurrence of cavities within it.  \pd{Later},
Liu \textit{et al.} \cite{liu2021molecular} 
identified various
regimes based on the We number, \pd{including} the splattering regime, and \pd{also suggested}
the `periphery-sucking' mechanism 
to explain
the thin middle and protruding edges of the expanding disk (e.g.,
see Figures~\ref{fig:1}e and~\ref{fig:1}j-k). Moreover, for
high kinetic energy of the droplets, various scenarios are possible,
namely, cavities in the bridge region, limited splattering, or
the so-called divergent splattering (\textit{e.g.} see
Figure~\ref{fig:1}i). 
}

\panos{Molecular dynamics \pd{(MD)} certainly offers specific advantages in investigating
phenomena such as droplet collisions.\cite{Wyatt1994,svanberg1998collision,Murad1999,Chun2011} 
In particular, it allows for 
capturing the microscopic mechanisms, which also evolve very
\pd{rapidly} 
in time. However, \pd{MD} studies have mostly been performed for systems
without surfactant. For example, Tugend et al.\cite{Tugend2025}
have carried out molecular dynamics simulations of large droplets
(up to $2\times10^7$ molecules) for a range of Weber and Reynolds
numbers. In their case, the coalescence, stable collision, holes, 
and shattering regimes have been observed, with the latter 
occurring after a critical We number and leading to the formation
of satellite structures. Some of these regimes have also
been observed previously by Greenspan and Heath.\cite{Greenspan1990}
The bouncing and coalescence regimes have also been 
investigated.\cite{Zhang2016,Jiang2016} as well as effects
of ambient pressure on nanodroplet collisions.\cite{zhang2019regimes,Zhang2021}
Finally, Liu et al.\cite{Liu2022} and Wang et al. \cite{Wang2024}
have investigated the coalescence of nanodroplets for a range of Weber
and Ohnesorge numbers, observing coalescence, stretching separation,
and shattering scenarios. \viooo{Also, oblique collisions of amorphous Lennard-Jones nanoparticles using molecular dynamics simulations as a function of collision velocity and impact parameter have been 
considered in the literature.\cite{nietiadi2022bouncing}.}
}

Qian \textit{et al.} \cite{qian1997regimes} conducted an experimental 
investigation of the collision of binary droplets
(water and hydrocarbon droplets) and constructed
a phase diagram based on the observed collision behaviour 
as a function of offset and Weber number.
Several distinct regimes were
identified, namely (I) coalescence after minor deformation, 
(II) bouncing, (III) coalescence after substantial deformation, 
(IV) coalescence followed by separation for near 
head-on collisions, and (V) coalescence followed
by separation for off-centre collisions. 
In particular, it has been found that regimes (II) and (III)
do not exist in droplets solely consisting of water.
Moreover, regimes IV and V lead to the formation of satellite 
droplets. \panos{Bouncing, which is due to a pressure build-up in the gap
between the droplet, was not observed in the case of pure water
droplets, which might be attributed to a 
higher surface tension and a lower viscosity in comparison with
hydrocarbon droplets. Pan \textit{et al.}\cite{pan2016controlling} studied the bouncing 
and coalescence of surfactant-laden aqueous droplets
by conducting experiments. Adding surfactant leads to 
a larger deformation of the surface, presumably, due to
a lower surface tension, and a higher probability of bouncing.
Experimental work by Krishnan et al.\cite{krishnan2015effects} on the collision
\pd{classified} 
five primary phenomenological outcomes, that is:
slow coalescence (SC), bouncing (B), fast coalescence (FC),
reflexive separation (RS), and stretching separation (SS).
}

\panos{The reflexive separation, which \pd{can be described as} temporary coalescence followed by breakup into two main droplets, has also been observed by Huang et al. \cite{huang2019pinching},
\pd{and} 
a larger number of satellite droplets is noted for higher We numbers. 
Munnannur et al. \cite{munnannur2007new} have developed a model for predicting 
collision outcomes, satellite formation, and post-collision characteristics (velocity and
droplet size) with predictions agreeing well with experimental results. 
Moreover, effects such as viscosity and surface tension have been discussed in 
a recent work by Pan et al. \cite{pan2009binary} and predictions for a critical
We number for satellite droplet formation have been made. 
In view of the role of viscosity and surface tension effects, which can be
expressed by the Ohnesorge number, various similarities between droplet collisions 
and the Plateau--Rayleigh instability could be drawn, \cite{Carnevale2023,Carnevale2024}
with a larger viscosity leading to less droplets in comparison with lower-viscosity
scenarios.\cite{wu2024evolution}
}

\panos{At this point, it is \pd{also} worth mentioning a few \pd{fundamental quantities} 
related to 
droplet coalescence before discussing our methods and results on droplets collisions.}
The rate at which the diameter of the bridge
$b$ (Fig.~\ref{fig:1} c) between droplets grows after their initial contact is crucial
for understanding the \panos{dynamics of the}
coalescence process. 
In coalescence studies, this growth can generally be characterised by two primary
fluid dynamics regimes: the viscous regime (VR) at early times 
and the inertial regime (IR) at later stages.\cite{Paulsen2014,paulsen2013approach} 
Additionally, recent MD simulations have
identified a third regime, known as the thermal regime (TR),\cite{perumanath2019droplet} 
which occurs during the very early stages of droplet coalescence when pinching takes 
place.\cite{perumanath2019droplet,arbabi2023bcoalescencePhysFluid,arbabi2023SoftMatter,arbabi2024SessilePhysFluid,Arbabi2023Spolymer}

In the viscous regime, the characteristic velocity, 
denoted as $\varv_{\rm v}$, can be expressed as $\gamma/\eta$, 
where $\gamma$ is the surface tension and $\eta$ the viscosity. 
Moreover, the Reynolds number can be expressed
as Re\,$=\rho V
b / \eta$, 
where $V$ 
is velocity, $b$ bridge \vioo{diameter} (Fig.~\ref{fig:1}c), 
which in the viscous regime becomes $\rho \gamma b / \eta^2$.
Since the bridge length is very small in this regime, 
viscous forces dominate regardless of the values of $\gamma$ 
and $\eta$, leading to Re\,$\ll1$. 
As coalescence progresses into the inertial regime (IR), 
the bridge velocity scales as
$V_i
\sim \sqrt[4]{{\gamma}/{\rho}}$.
The crossover between the viscous 
and inertial regimes is expected to occur when 
${\rm Re} \sim 1$. 
The characteristic viscous time scale  of
$t_{\rm v} = {\eta R_0 / \gamma}$
and the characteristic inertial time scale of
$t_{\rm i} = \sqrt{\rho R_0^3 / \gamma}$ have
been suggested by a lattice Boltzmann 
study,\cite{gross2013viscous} where
$R_0=D_0/2$ \vioo{(Fig.~\ref{fig:1}a)}.

In the VR, where intermolecular forces predominantly 
drive the coalescence process, the bridge diameter
for coalescence of freely suspended droplets has been proposed
to scale linearly with time, expressed as $b \propto t$,
with some suggesting logarithmic corrections, that is
$b \propto t \ln t$.\cite{duchemin2003inviscid,eggers1999coalescence}
For the IR, a power-law scaling has been proposed 
for the bridge diameter, specifically $b \propto \sqrt{t}$.\cite{duchemin2003inviscid,eggers1999coalescence} 
Experimental studies on the coalescence of 
water droplets support this scaling behaviour.\cite{eggers1999coalescence,aarts2005hydrodynamics,thoroddsen2005coalescence,gross2013viscous,sprittles2012coalescence}
In addition, it has been suggested 
that the inertia of the droplets cannot be ignored 
during the initial stage of coalescence. This 
initial stage is then  better described as an 
inertially limited viscous (ILV) regime, 
where a linear scaling of the bridge 
dimension with time has been proposed.
\cite{paulsen2012inexorable,paulsen2013approach} 
\panos{However, the existence of the ILV regime
seems to remain questionable with 
Eggers et al. arguing that the bridge
region remains purely viscous. \cite{eggers2024coalescence} }

All-atom molecular dynamics simulations of 
two-dimensional (cylindrical)
pure water droplets have revealed multiple liquid bridges 
forming on the droplet surfaces and connecting them, which 
are highly affected by thermal fluctuations at the 
molecular level and indicate the so-called thermal regime at 
the beginning of the coalescence
process.\cite{perumanath2019droplet} 
In this case, once the bridge \vioo{length} exceeds a thermal length scale, 
estimated 
as
$l_T \approx {\left ( k_BT/\gamma \right) }^{1/4} R_{0}^{1/2}$, 
the system transitions to a hydrodynamic regime. Here,
$\gamma$ is surface tension,
$T$ temperature, and $k_B$ Boltzmann's constant.
We have previously demonstrated the presence of the thermal regime
followed by the inertial regime for both 
freely suspended\cite{arbabi2023SoftMatter,arbabi2023bcoalescencePhysFluid}
and sessile droplets \cite{arbabi2024SessilePhysFluid}, 
including both the case of pure water droplets as well
as that with surfactant-laden droplets.
A summary of bridge growth scalings in
the case of zero-velocity droplet-coalescence found
in our earlier studies based on the same
MD model are reported in Table.~\ref{tbl:table1}.

\begin{table*}[bt!]
\centering
    \caption{Summary of bridge growth scaling
    within the inertial regime in freely suspended and sessile droplets from MD studies based on the same model.
\cite{arbabi2023bcoalescencePhysFluid,arbabi2023SoftMatter,arbabi2024SessilePhysFluid,Arbabi2023Spolymer}.}
    \label{tbl:table1}
    \begin{tabular*}{0.7\textwidth}{@{\extracolsep{\fill}}ll}
    \hline
    \textrm{System} 
    & \textrm{Bridge dimension growth ($b$)}\\
    \hline
    
    \small{Suspended water and surfactant-laden droplets \cite{arbabi2023bcoalescencePhysFluid,arbabi2023SoftMatter}}    & $b \sim t^{0.5-0.6}$ \\

    \small{Sessile water and surfactant-laden droplets} ($\theta_s \geq 90^{\circ}$)  \small{\cite{arbabi2024SessilePhysFluid}}   & $b \sim t^{0.5-0.6}$ \\

    \small{Sessile water and surfactant-laden droplets} ($\theta_s < 90^{\circ}$) \small{\cite{arbabi2024SessilePhysFluid}}  & $b \sim t^{0.6-0.8}$ \\

    \small{Sessile polymer droplets}  ($\theta_s > 90^{\circ}$) \cite{Arbabi2023Spolymer}   & $b \sim t^{0.28-0.38}$     \\
  \small{Sessile polymer droplets} ($\theta_s < 90^{\circ}$) \cite{Arbabi2023Spolymer}    & $b \sim t^{0.29-0.45}$   \\ 
   \hline

    \end{tabular*}
\end{table*}

\panos{Despite progress in the study of droplet collision, especially
from the point of view of molecular dynamics
simulations of large systems,\cite{Tugend2025}
which is suitable for describing
the microscopic details of this fast process, the role of
surfactant has largely remained unexplored.}
To fill this gap, this study builds on previous work on the 
coalescence of pure and surfactant-laden water 
droplets\cite{arbabi2023SoftMatter,arbabi2023bcoalescencePhysFluid}
and explores the head-on and off-centre collision for 
water and surfactant-laden droplets
by means of molecular dynamics (MD) simulation based on
a coarse-grained force-field. Our findings suggest that the presence of 
surfactants significantly affect the dynamics
of collisions and, in particular, the head-on
collisions, with significant differences appearing
in comparison with the droplet coalescence at
zero velocity.\cite{arbabi2023SoftMatter,arbabi2023bcoalescencePhysFluid}

The paper is organised as follows: 
\panos{In the following section}, we detail our model 
and methodology, while in Section~\ref{results} we 
present the analysis and classification of the results of head-on
and off-centre collisions with details provided on the oscillation states and final
states of the droplets. Finally, Section~\ref{conclusions}
offers conclusions drawn from this study.

\section{Model and Methods}
\label{model}

\begin{figure}[!htbp]
\centering
\includegraphics[width=\columnwidth]{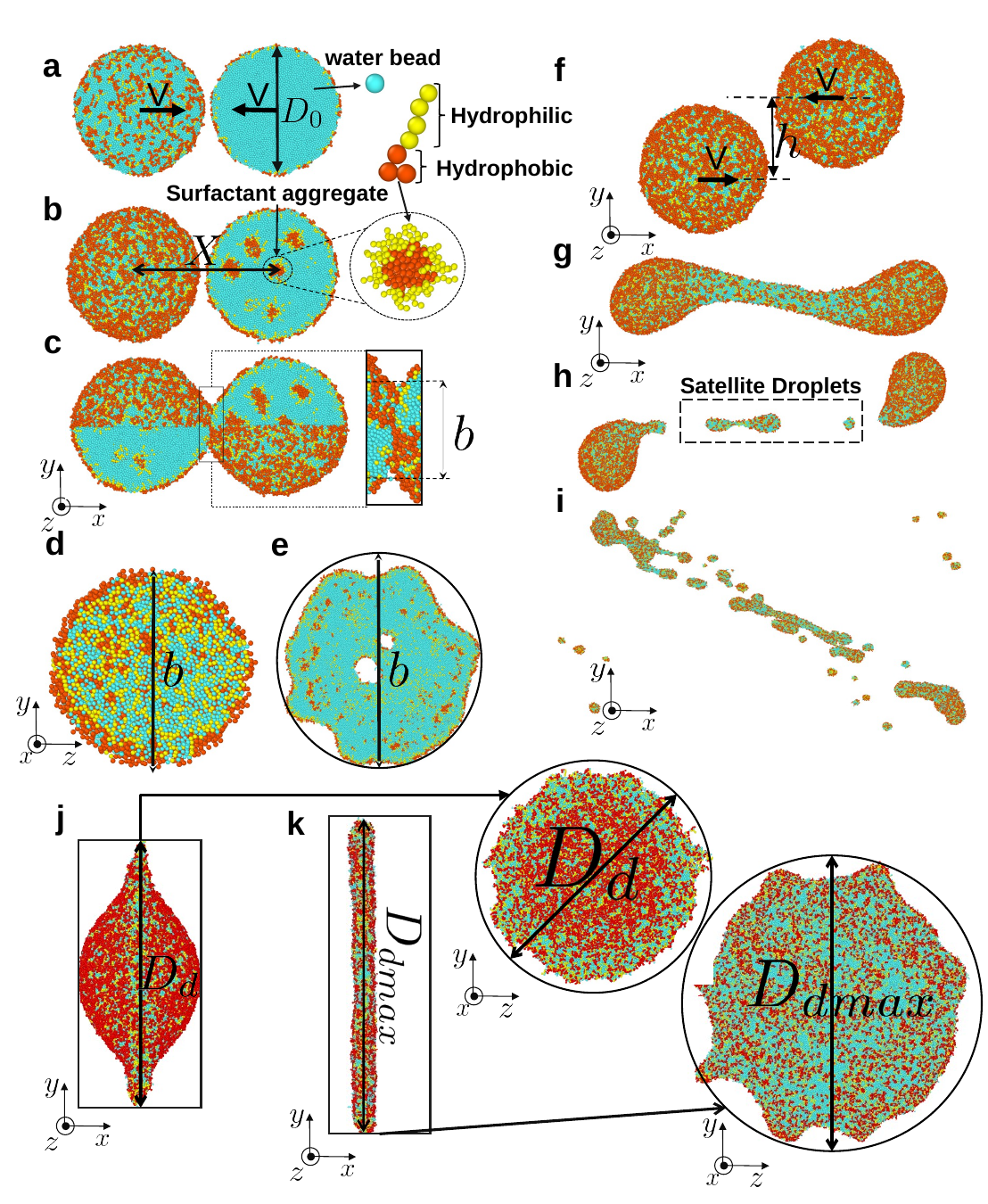}
\caption{\label{fig:1} 
Various stages of droplet collision for a number of 
scenarios 
differing in surfactant concentration and velocities. 
Scales are adjusted
in the figure, so that the structural characteristics of the droplets
and surfactant molecules are visible. The left column shows a
head-on collision, while the right column an off-set one. 
Panels a-b show initial configurations before a head-on collision, below (a) and
above (b) the critical aggregation concentration.
In the latter case, a micelle and one of the participating surfactant molecules
are shown. 
(c) A snapshot of an above CAC case early in the collision.  
The bridge \vioo{diameter} is indicated as $b$.
(d) Snapshot of a disk that forms
as a result of 
a higher velocity collision, where the
point of contact can keep expanding until the liquid forms
a disk-like structure.
This ultimately has a larger diameter ($D_d$, \panos{panel j}) than the initial \vioo{diameter of the droplets} ($D_0$) in the direction
perpendicular to the collision axis.  
At a high enough velocity (e),
cavities (holes) may appear after reaching
$D_{dmax}$ \vioo{(panel k)}, which is 
the maximum diameter that the disk reaches during the collision
($V=2.2415~\sigma / \tau$, $t=152~\tau$).
(f) Configuration of the system before
an offset collision with a vertical offset
$h$. Various snapshots show different 
scenarios, being: 
(g) the stretching after the collision 
(velocity $V=1.3449~\sigma / \tau$, $h = 0.5~D_0$, $t=150~\tau$),
(h) the formation of satellite droplets ($V=1.3449~\sigma / \tau$, $h = 0.5~D_0$, $t=290~\tau$), 
and 
(i) splattering, which happens in the
case of high-velocity collisions
($V=2.9146~\sigma / \tau$, $h = 0.3~D_0$, $t=140~\tau$).
(j, k) Bridge and disk formation
during the head-on collision.
When the bridge diameter 
 equals the droplets' diameter (\vioo{$b=D_0$}), 
 the disk between the droplets continues to 
 grow if the kinetic energy (from the initial collision velocity) 
 wins over
 the damping. For pure water and below CAC surfactant-laden droplets, this occurs when $V\simeq 0.67\,~\sigma / \tau$, and $V\simeq 0.89 \,~\sigma / \tau$, \panos{respectively}. We then denote the disk diameter
 as $D_d$, with side views provided under (j) and (k).
}
\end{figure}

\panos{Before providing more technical detail on the 
simulations, we 
\pd{first give a brief} overview of the
different scenarios observed during our simulations.}
\vio{Droplets can collide head-on or off-centre  
(Figure~\ref{fig:1}). 
A head-on or nearly head-on (\panos{paraxial,}small offset, $h$) 
collision leads to the creation of a disk with diameter
$D_d$ as it goes through the bridge formation stage and coalescence, as shown in Figures~\ref{fig:1}j--k.
In this case, $D_d$ is greater than the diameter $D_0$ of the
initial droplet. This disk
expands (Fig.~\ref{fig:1}d and Figure~\ref{fig:1}j--k),
and, moreover, holes (cavities) will form within the disk 
(Figure~\ref{fig:1}e), which might be attributed to the fact
that the kinetic energy during the disk expansion \pd{wins over} 
the viscous dissipation. In Figure~\ref{fig:1}c, the bridge, and, in Figures~\ref{fig:1}j and k, the disk  are depicted to better highlight their differences.
In the case of off-centre collisions     
(Figure~\ref{fig:1}f-i), significant stretching of the 
liquid (Figure~\ref{fig:1}g) and the creation of satellite
droplets are observed (Figure~\ref{fig:1}h). 
In addition, at higher collision velocities, 
both the head-on and off-centre droplets collision can lead to 
splattering (Figure~\ref{fig:1}i).
}

\panos{At a more technical level,} we conduct our research using 
MD simulations based on the SAFT  (Statistical Associating Fluid Theory) $\gamma$-Mie force-field.\cite{chapman1989saft,muller2001molecular,Avendano2011,Avendano2013,sergi2012coarse,muller2014force} 
In the case of surfactant-laden droplets, the
SAFT-$\gamma$ Mie theory force-field,\cite{Lafitte2013} 
has been accurate in reproducing key properties
of water--surfactant systems, such as phase behaviour, 
contact angles of droplets, and surface 
tension.\cite{Theodorakis2015modelling, Theodorakis2019molecular, lobanova2014development, lobanova2016saft, morgado2016saft}
Moreover, as a coarse-grained (CG) force-field, 
it enables
the simulation of relatively large droplets, 
which in turn allows for a careful examination of
surfactant-related mechanisms and relevant properties
as has been done in the case of the coalescence of surfactant-laden
droplets.\cite{arbabi2023bcoalescencePhysFluid,arbabi2023SoftMatter,arbabi2024SessilePhysFluid}
Other applications of droplet-related phenomena with this 
force-field included
study of the superspreading of surfactant-laden 
droplets.\cite{Theodorakis2015modelling,Theodorakis2019molecular,Theodorakis2015Langmuir,Theodorakis2014,Theodorakis2019} 

In the SAFT $\gamma$-Mie force-field, interactions between 
different coarse-grained (CG) beads within a distance
smaller than $r_{\rm c}$ are described
via the Mie potential
\begin{equation}
\label{equation_mie}
    U(r_{\rm ij}) = C\epsilon_{\rm ij} \left[ \left({\frac{\sigma_{\rm ij}}{r_{\rm ij}}}\right)^{\lambda_{\rm ij}^{\rm r}} - \left({\frac{\sigma_{\rm ij}}{r_{\rm ij}}}\right)^{\lambda_{\rm ij}^{\rm a}}\right],
     r_{\rm ij} \leq r_{\rm c},
    \end{equation}
where
\begin{equation}
    C = \left(\frac{\lambda_{\rm ij}^{\rm r}}{\lambda_{\rm ij}^{\rm r} - \lambda_{\rm ij}^{\rm a}}\right){\left( \frac{\lambda_{\rm ij}^{\rm r}}{\lambda_{\rm ij}^{\rm a}}\right)}^{\frac{\lambda_{\rm ij}^{\rm a}}{\lambda_{\rm ij}^{\rm r} - \lambda_{\rm ij}^{\rm a}}}.
\end{equation}
i and j are the bead types, $\sigma_{\rm ij}$ indicates the 
effective bead size,
and $\epsilon_{\rm ij}$ is the interaction strength
between any beads of type i and j. 
$\lambda_{\rm ij}^a=6$ and $\lambda_{\rm ij}^r$ are 
Mie potential parameters,
while $r_{\rm ij}$ is the distance between two CG beads.
The chosen units are for the length, $\sigma$, energy, $\epsilon$, mass, $m$, and
time $\tau$ corresponding to
real units as follows: 
$\sigma = 0.43635$~nm, $\epsilon / k_B = 492$~K, $m=44.0521$~amu
and $\tau = \sigma{(m/{\epsilon})}^{0.5} = 1.4062$~ps. 
The simulations were performed at room temperature ($T=25^{~\circ}$C),
which corresponds to $T=0.6057~\epsilon/k_B$ in simulation units. 
A universal cutoff $r_c = 4.583~\sigma$ was 
applied to all nonbonded (Mie) interactions.

\begin{figure}[bt!]
\centering
\includegraphics[width=\columnwidth]{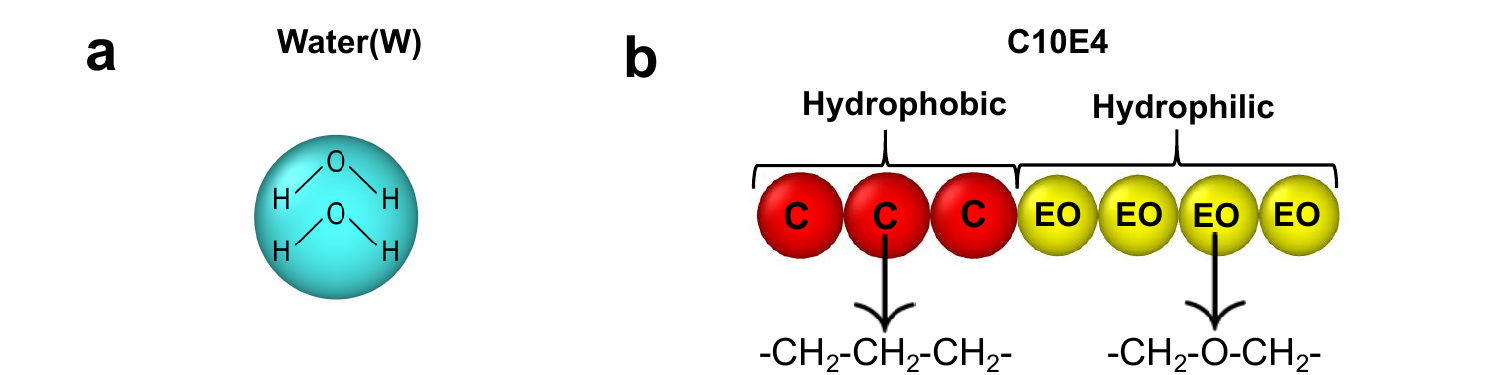}
\caption{\label{fig:3} Coarse-grained representation 
of two water molecules (a) and a surfactant molecule (b). The
hydrophobic beads of the surfactant are shown in red, while 
the hydrophilic ones are in
yellow.
}
\end{figure}

A surfactant of type CiEj (Figs~\ref{fig:3}b) is considered in this study, \textit{i.e.} C10E4.
In general, in the case of CiEj surfactants, a hydrophobic alkane
CG `C' bead represents a $\rm -CH_2-CH_2-CH_2-$ 
group of atoms,
while a hydrophilic CG `EO' bead represents an
oxyethylene group $\rm -CH_2-O-CH_2$.
A water CG `W' bead corresponds 
to two water molecules (Fig.~\ref{fig:3}a).
The non-bonded interaction parameters between
the above chemical groups  
are reported in Table~\ref{tbl:table2},
while the mass of each CG bead is documented 
in Table \ref{tbl:table3}.

\begin{table}[bt!]
\small
    \caption{Summary of Mie interaction parameters (Eq.~\ref{equation_mie}). $\lambda_{\rm ij}^{\rm a} = 6$. }
    \label{tbl:table2}
    \begin{tabular*}{0.48\textwidth}{@{\extracolsep{\fill}}lccr}
    \hline
    \textrm{i--j}&
    \textrm{$\sigma_{\rm ij}~[\sigma]$}&\textrm{$\epsilon_{\rm ij}~[\epsilon/k_B]$}&\textrm{$\lambda_{\rm ij}^{\rm r}$} \\
    \hline
    W--W & 0.8584 & 0.8129 & 8.00\\
    W--C & 0.9292 & 0.5081 & 10.75 \\
    W--EO & 0.8946 & 0.9756 & 11.94 \\ 
    C--C & 1.0000 & 0.7000 & 15.00 \\
    C--EO & 0.9653 & 0.7154 & 16.86\\
    EO--EO & 0.9307 & 0.8067 & 19.00\\
        \hline
    \end{tabular*}
\end{table}
    
\begin{table}[bt!]
\small
    \caption{Mass of the CG beads.}
    \label{tbl:table3}
    \begin{tabular*}{0.48\textwidth}{@{\extracolsep{\fill}}cc}
    \hline
    \textrm{Bead Type}&
    \textrm{Mass~[m]} \\ 
    \hline
    W & 0.8179 \\
    C & 0.9552  \\
    EO & 1.0000\\ 
        \hline
    \end{tabular*}
\end{table}

In the case of surfactant
chains, a bond potential is required to tether consequent beads along the 
chain, which
in the case of this model is 
\begin{equation}
\label{equation_bonded1}
    V_{\rm bond}(r_{\rm ij}) = 0.5k(r_{\rm ij}-\sigma_{\rm ij})^2,
\end{equation}
where the harmonic constant is $k = 295.33$~$\epsilon/\sigma^2$.
Moreover, EO  beads experience a harmonic angle potential,
\begin{equation}
\label{equation_bonded2}
    V_\theta(\theta_{\rm ijk}) = 0.5k_\theta(\theta_{\rm ijk}- \theta_0)^2,
\end{equation}
where $\theta_{\rm ijk}$ is the angle formed by three consecutive
beads i, j, and k, (regardless of bead type), 
$k_\theta = 4.32$~$\epsilon/$rad$^2$, 
and $\theta_0 = 2.75$~rad is the equilibrium angle.
Additional details about the model can be found in previous studies. \cite{Theodorakis2015Langmuir,Theodorakis2015modelling,Theodorakis2019}

To prepare the initial configuration for each system, as done in earlier studies,\cite{arbabi2023bcoalescencePhysFluid,arbabi2023SoftMatter}
individual droplets were first equilibrated within the NVT ensemble using the 
Nosé-–Hoover thermostat through the LAMMPS package\cite{LAMMPS}
with an integration time step of $\delta t = 0.005~\tau$. 
Each initial droplet contained $10^5$ beads in the simulations, 
with approximately 5\% evaporating into the gas phase. 
The droplet diameters were $\sim53~\sigma$ (approximately 23~nm), 
consistent with several previous studies.\cite{perumanath2019droplet,arbabi2023bcoalescencePhysFluid,arbabi2023SoftMatter} 
\viooo{Additional information and the database required to reproduce the data are provided in the ESI section titled \textit{Simulation Parameters and Data Availability}.}
Attention was given not only to monitoring the system's energy but also 
to ensure that the surfactant clusters reached dynamic equilibrium, 
allowing each cluster to diffuse a distance many times its size.
Once the individual droplets were equilibrated, 
the two droplets, along with the surrounding gas, 
were positioned next to each other as depicted in Fig.~\ref{fig:1}a in the case of a 
head-on collision or with the desired offset, $h$ (Fig.~\ref{fig:1}f),
in the case of off-centre collisions, preserving in the two-droplet system the same number of particles per volume as in the 
single-droplet simulations. At this stage, the desired offset ($h$) 
and centre-of-mass velocities ($V$) are assigned to each droplet, 
and the collision simulation is performed in the NVE ensemble with a time step 
of $\delta t = 0.001~\tau$. \panos{The smaller time step here ensures
that the MD simulations remain stable during the integration of the
equations of motion for the particles for the highest We numbers of this study.}

The final size of the simulation box was selected to 
be sufficiently large to prevent \pd{any} interactions
between mirror images of the droplets due to the presence of periodic boundary conditions in all directions,
and moreover, to ensure that the box is large enough
to be able to observe the elongated structures in the offset
collisions as shown in Figure~\ref{fig:1}g--h.
We simulate surfactant-laden droplets 
below and above the critical aggregation concentration (CAC) as well as
pure water droplets for 
    comparison.
Figure~\ref{fig:1}a illustrates a typical initial snapshot
for cases below CAC, while Figure~\ref{fig:1}b shows a
case above CAC. 
A summary of the mean values of various properties for
our systems with surfactant is given in Table~\ref{tbl:table4}. 
\panos{Information on systems that have been
studied in the literature (e.g. We numbers, drop-size
ratios and sizes, ambient medium, etc.) can be found
in Ref.~\citen{munnannur2007new}, while the range of
parameters considered in recent molecular dynamics simulations
and the span of We numbers for which bouncing, coalescence,
stable collision, holes, and shattering occur 
can be found in Ref.~\citen{Tugend2025}.}
\vio{Finally, we calculated the surface tension of the water droplet to be approximately $72\,\mathrm{mN/m}$, while that of the surfactant-laden droplets discussed above CAC is approximately $29\,\mathrm{mN/m}$.
\vioo{Considering these surface tension values, the Weber numbers for water droplets in this system range from near 0, corresponding to coalescence, up to approximately 184, this upper limit occurs at a velocity of $V\simeq 2.4662 \,~\sigma / \tau$. For surfactant-laden droplets, the Weber numbers can be even higher due to the reduced surface tension, which enhances the influence of inertial forces relative to surface tension.} To examine \pd{comparable, matching} 
cases for both pure water droplets and droplets laden with surfactant, we chose here to present our results based on the velocity, instead
of dimensionless numbers.} 
\panos{Moreover, determining exact values for the viscosity
of complex systems (e.g. those containing surfactants)
in molecular dynamics
simulations with potentials of hard-core interactions can be challenging. For this reason, a discussion of the results
in the context of Reynolds numbers, as is done
in other studies,\cite{Tugend2025} is omitted here. }

\begin{table}[bt!]
  \caption{\small Properties of individual droplets (equilibrium) }
  \label{tbl:table4}
  \footnotesize  
  \begin{tabular*}{0.45\textwidth}{@{\extracolsep{\fill}}cccc}
    \hline
    Concentration (wt\%) & Diameter ($\sigma$) & Water Beads* & Surfactant Molecules\\
    \hline
    \textbf{Water} & & & \\
    - & 52.5 & 95882 & - \\
    
    \textbf{C10E4} & & & \\
    6.25 & 53.1 & 90466.1 & 714  \\
    35.48 & 54.1 & 65746.7 & 4286 \\
    CAC $\approx$ 7.5~wt\% & & & \\
    \hline 
  \end{tabular*}
  \normalsize  
  *Indicates the average number of water beads.
\end{table}

\section{Results and Discussion}
\label{results}

\subsection{Head-On Collision}

The first part of this study focuses on head-on collisions, where
two droplets collide face-to-face with zero offset 
(Fig.\ \ref{fig:1}a--b). 
Considering that the gravitational potential energy 
 and the dissipation away from the bridge,
are not relevant (the droplet size is smaller than
the capillary length of water, \panos{for example,
see Table~\ref{tbl:table4} for the droplet sizes
in this study}), the approximate energy balance of the relevant
elements of the system can be expressed as follows:
\begin{equation}
\label{energy_balance}
    E_{k1} + E_{S1} = E_{k2} + E_{S2} + W.
    \end{equation}
Here, $E_{k1}$ and $E_{k2}$ are the kinetic energy of each droplet
and $E_{s1}$ and $E_{s2}$ are the surface energies
of each droplet that can be estimated
as follows:\cite{zhang2019regimes,li2017spreading}
\begin{equation}
\label{energy_surface}
    E_{S1} = 2\pi{D_0}^2\gamma,
    \end{equation}
\begin{equation}
\label{energy_kinetick}
    E_{k1} = \frac{\pi {D_0}^3 \rho V^2}{6},
    \end{equation}
where $D_0$ is the diameter of the droplets. The quantity $W$
is the viscous dissipation from the initial state 
to a maximum spreading state
which is defined
as follows\cite{zhang2019regimes,li2017spreading}:
\begin{equation}
\label{dissipation_energy}
     W = 2 \int_{0}^{t_c} dt\int_{\Omega} \mu V D_0^2 (\beta^5 - 1) = \frac{3 \pi}{80} \mu V D_0^2 (\beta^5 - 1),
    \end{equation}
where $\beta$ is the maximum spreading factor
 \begin{equation}
\label{beta}
     \beta = \frac{D_{dmax}}{D_0}.
    \end{equation}
Here, $D_{dmax}$ is the maximum spreading diameter, which is the diameter of the disk in our case.
Moreover, Zhang \textit{et al.} \cite{zhang2019regimes} obtained the following 
relation between maximum spreading factor, Re number, and We number:

\begin{equation}
\label{Beta-We-Re}
    \frac{3}{80} ({\beta}^5 - 1)+\frac{Re}{We}({\beta} ^2 + \frac{1}{3\beta} - 2)-\frac{Re}{6} = 0
\end{equation}

\begin{figure}[bt!]
\includegraphics[width=1.0\columnwidth]{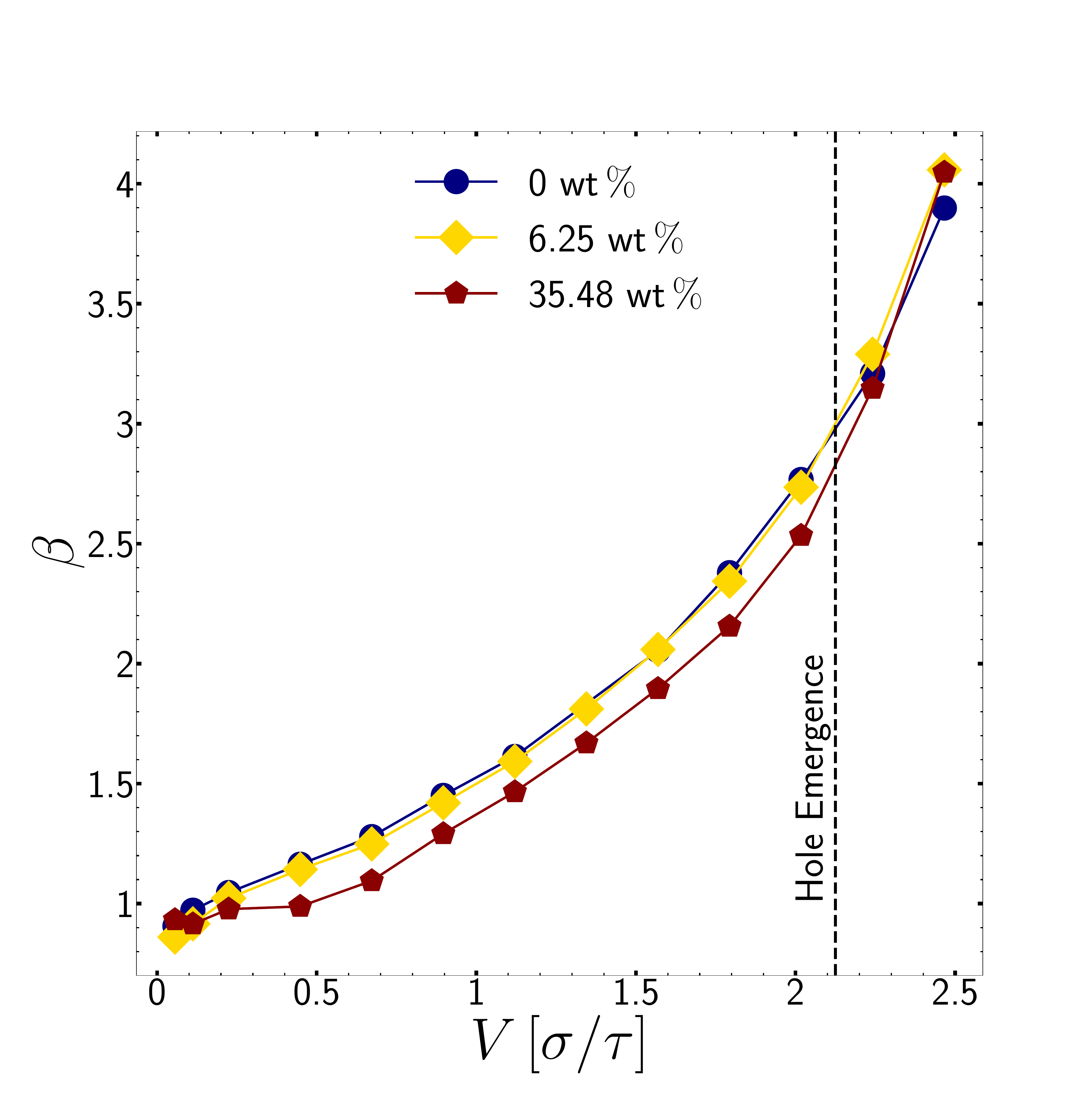}
\caption{\label{fig:fig4} 
Comparison of $\beta=D_{dmax}/D_0$ parameter that 
quantifies the relative size of the disk for 
different collision velocities during head-on collisions. 
In all cases, when the velocity was 
$V \simeq 2.24~\sigma / \tau$,
cavities
were observed in the bridge.  
}
\end{figure}

Figure~\ref{fig:fig4} shows the evolution of the spreading factor, $\beta$, 
during the collision of water droplets and surfactant-laden droplets.
Pure water droplets and droplets with surfactant
concentration below CAC have a very similar trend, while, above CAC 
(CAC$\approx 7.5$~wt\%) $\beta$ obtains lower 
values at similar collision velocities. 
Moreover, the maximum ratio for which cavities appear
$\beta=3.0\pm0.2$ that can be estimated from
this study
 is reasonably close to 
the value of 2.66 reported in Ref.~\citen{zhang2019regimes}, though slightly higher.

\subsubsection{Bridge Dynamics}
During the first part of the collision, 
a bridge of diameter $b<D_0$ forms between the two droplets 
(Figure~\ref{fig:1}c). In our previous studies
\cite{arbabi2023bcoalescencePhysFluid,arbabi2023SoftMatter} 
related to the coalescence of freely suspended droplets, 
we have demonstrated that the  bridge
initially appears as a small  overlap with fluctuating diameter $b$
within the thermal regime, followed by
power-law growth of the diameter in the inertial regime.
In case of droplet collisions, 
we observe the emergence of a linear regime.
This linear regime is more pronounced in the
collision process for larger velocities, when
the collision time becomes smaller.

\begin{figure}[bt!]
\includegraphics[width=1.0\columnwidth]{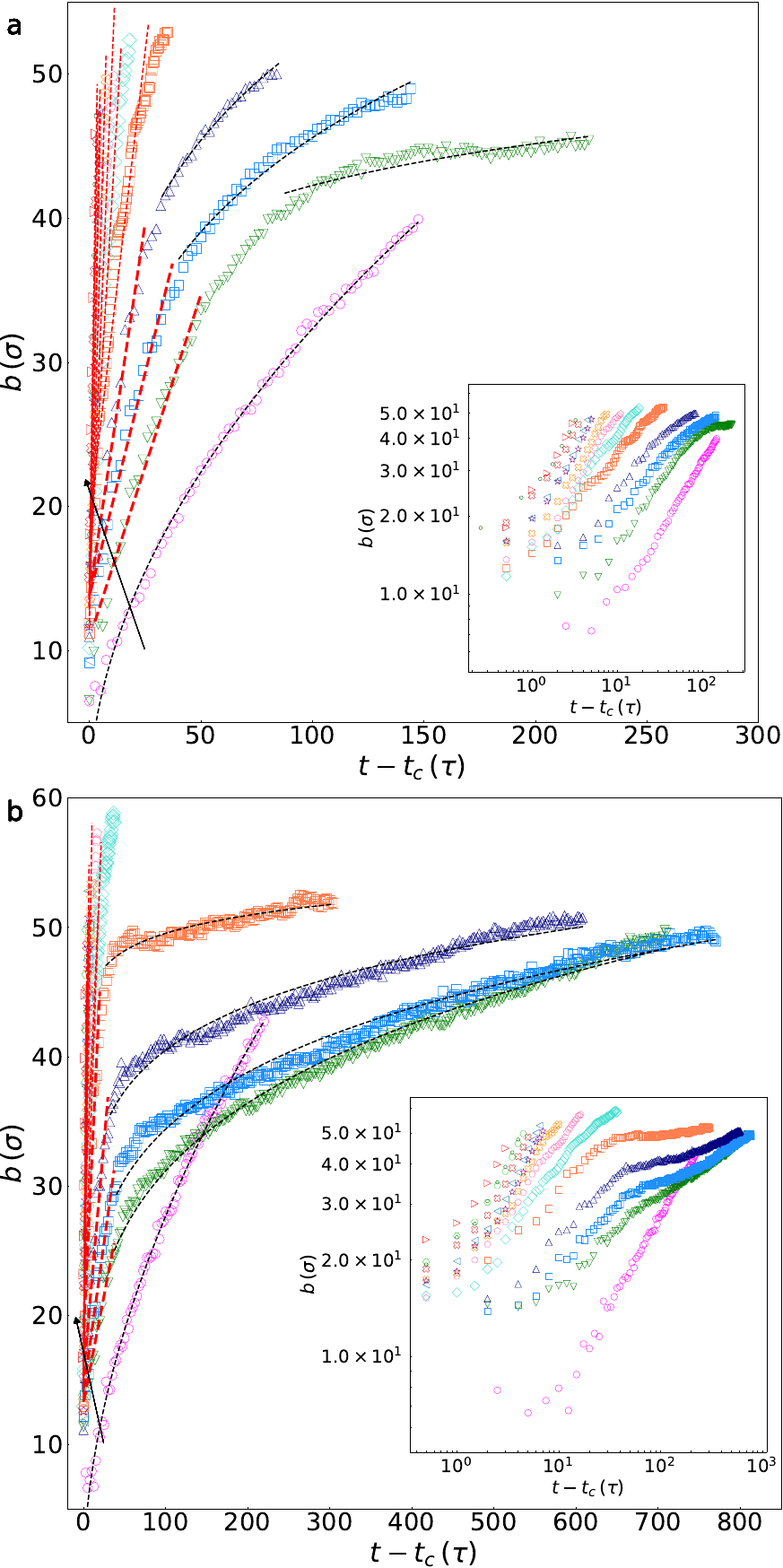}
\caption{\label{fig:fig5} Bridge dynamics during head-on collisions of (a) water and (b) 
surfactant-laden (above CAC) droplets at different velocities. In each case, 
the linear fit is represented by a red dashed line with the 
fitting parameter $m_v$, while the power-law fit is 
illustrated by a black dashed line with the 
fitting parameter $\alpha_i$. 
The black arrow  indicates the progression from lower to higher velocities \pd{in the range 0 to 2.4662$\sigma/\tau$}, 
and the details of the fitting parameters
are provided in \panos{Tables~S1 and S2 of the Supporting Information}. 
In the plots, $t_c$ indicates that all measurements are taken from the time ($t_c$)
droplets make contact.
}
\end{figure}

Figure~\ref{fig:fig5} presents the dynamics of the bridge diameter 
in the case of pure water (Figure~\ref{fig:fig5}a) and 
surfactant-laden (Figure~\ref{fig:fig5}b)
droplets (above CAC), respectively. In each panel,
the linear regime is fitted by red dashed lines with slopes $m_v$ reported in Tables~\panos{S1} and
\panos{S2 of the Supporting Information}, for
Figures~\ref{fig:fig5}a and ~\ref{fig:fig5}b, respectively.
The inertial regime is fitted via a black dashed line and 
the power-law exponents are reported as $\alpha_i$
in the same tables.

At medium velocities of droplet collision 
(from $0.056~\sigma / \tau$ to $0.2242~\sigma / \tau$ 
for water and from $0.056~\sigma/ \tau$ to $0.4483~\sigma / \tau$
for surfactant-laden 
droplets above CAC in our data) two regimes exist. Both the linear and 
power-law growth of bridge size are steeper in the case of 
pure water droplets.
At higher velocities (above $0.2242 ~\sigma / \tau$ for water and above
$0.4486~\sigma / \tau$ for surfactant-laden droplets  in our data), 
the power-law regime disappears. 
Here, 
the kinetic energy is large and the 
impact energy cannot be completely damped by viscous dissipation
and surface energy. For this reason, the power law growth characteristic of slow coalescence does not appear.
Surfactant-laden droplets have higher viscosity and smaller
$\beta$ (Figure~\ref{fig:fig4}) than water droplets,
which implies 
higher rates of viscous dissipation of energy that lead to smaller expansion of the bridge between the droplets.
It can be observed that surfactant-laden droplets were able to
maintain coalescence behaviour up to higher velocities, where $V \simeq 0.4483~ \sigma / \tau$,
while water droplets can sustain the coalescence regime only 
up to 
$V \simeq 0.2242  ~\sigma / \tau$. 
\vio{It is worth mentioning that, in the case of surfactant-laden droplets (Figure~\ref{fig:fig5}b), the values \pd{of $b$} for the coalescence case ($V=0~\sigma / \tau$) initially increase more slowly compared to the other data sets. However, there is a \pd{crossing} 
with the collision cases at velocities between $0.056~\sigma / \tau$ and $0.1121~\sigma / \tau$, indicating that the 
bridge \pd{growth after coalescence} can
eventually \pd{turns out} 
faster \pd{than after a collision}. This phenomenon might be explained by the fact that, when droplets are more viscous, the kinetic energy during low-velocity collisions is significantly dampened. As a result, the energy \pd{becomes} insufficient to \blu{continue} 
the bridge growth, and \pd{the} 
effect \pd{ends up} actually delaying    it. }

\begin{figure}[bt!]
\includegraphics[width=1.0\columnwidth]{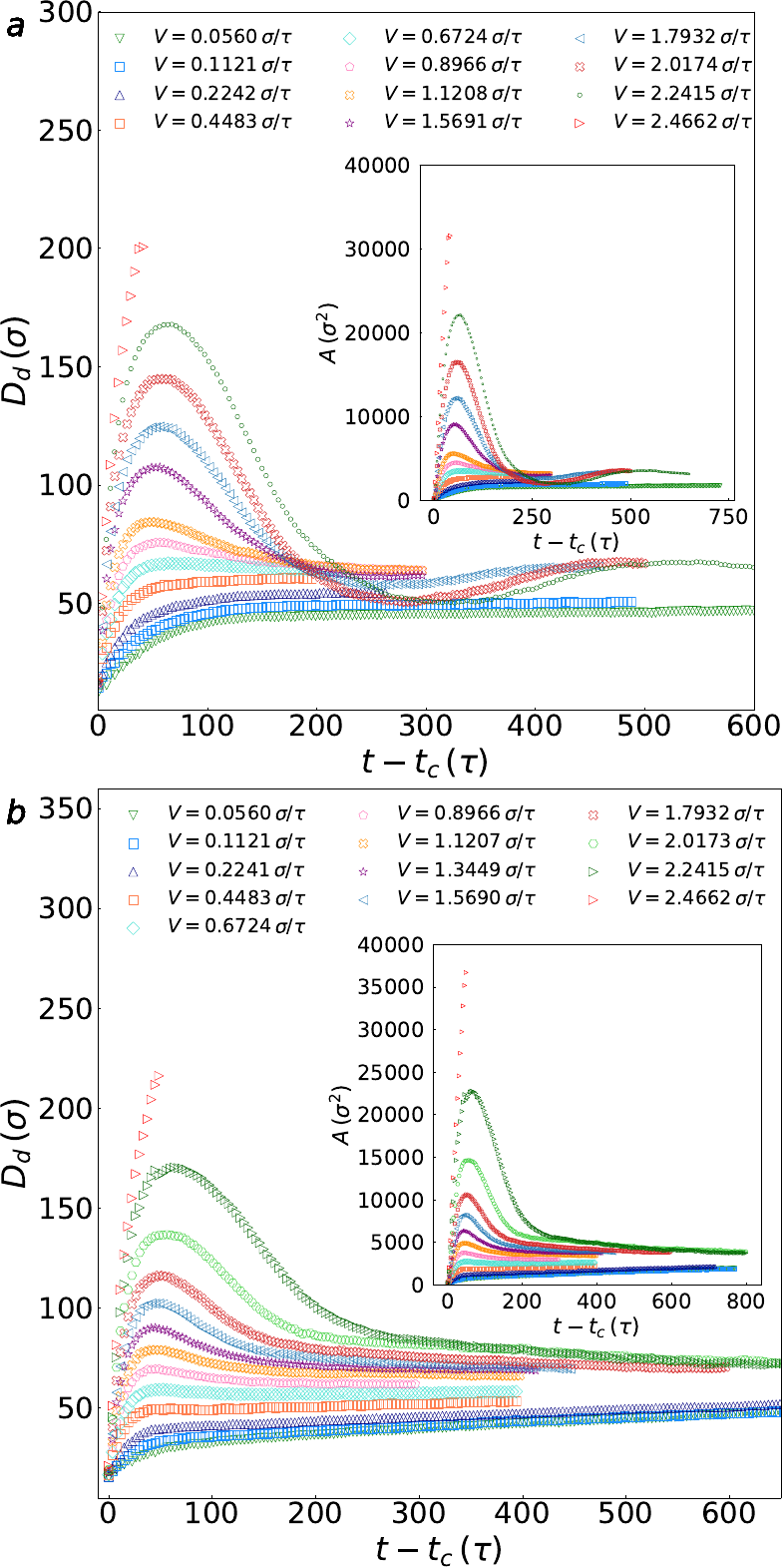}
\caption{\label{fig:fig6}
Disk dynamics during head-on collisions of (a) water 
droplets and (b) surfactant-laden 
droplets (above CAC) at different velocities.
The area of the disk is shown in the inset of the figure. }
\end{figure}

\subsubsection{Disk Dynamics}

We further analyse the disk diameter during the collision process (Fig.~\ref{fig:1}d,e,j,k).
In the previous section, we considered the bridge up to the moment 
when its diameter $b$ exceeds the initial droplet diameter $D_0$.
However, in a fast enough collision, the point of contact can keep
expanding until the liquid forms a disk-like structure that has a 
larger \vioo{diameter} than the initial droplets, but usually a shorter
thickness in the collision direction than $D_0$ (the droplet's initial diameter is around $50~\sigma$).
This disk can expand until the viscous dissipation consumes
the initial kinetic energy of the expanding disk.

The average disk diameter ($D_d$ in Figure~\ref{fig:1}j) 
is plotted over 
time for water droplet collisions (Figure~\ref{fig:fig6}a) and 
surfactant-laden droplet collisions above CAC (Figure~\ref{fig:fig6}b). 
The most visible
difference between these cases is that, 
for water droplets at high velocities (runs with $V$ above $1.5691 ~\sigma / \tau$), 
the disk \vioo{diameter} starts to \vio{oscillate} 
after reaching its maximum \vioo{diameter}. 
In contrast, for surfactant-laden droplets (above CAC), 
there is no fluctuation; instead, the disk shrinks monotonically
after  reaching its maximum \vioo{diameter}. 
The absence of fluctuations in the case of surfactant-laden droplets may 
be due to the greater viscous dissipation, 
which more efficiently absorbs the initial kinetic energy.

\begin{figure}[bt!]
\includegraphics[width=1.0\columnwidth]{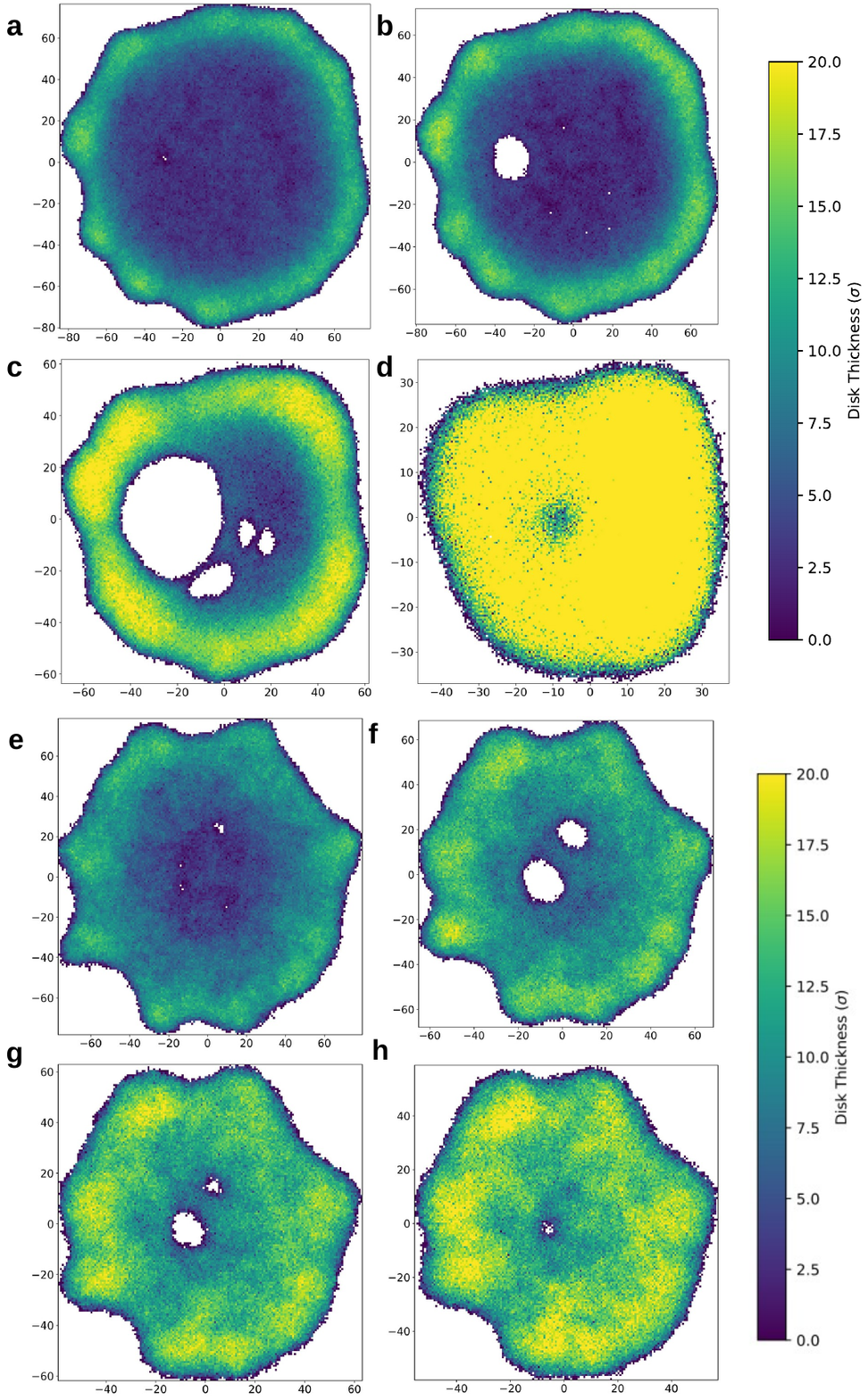}
\caption{\label{fig:fig7}  \vio{The \pd{plots show a} comparison \pd{of disks for} 
(a-d) pure water and (e-h) surfactant-laden droplets above the CAC. The plot illustrates the thickness of the disk, showing that in the case of pure water, there is a stronger flow towards the rim of the disk, which leads to a thinner middle region and larger holes. In contrast, for surfactant-laden droplets, there is a weaker flow, resulting in a thicker middle region and smaller holes.
The view of the disk is oriented along the $x$ axis. For pure water, the time sequences are as follows: (a) $ t \simeq 90.75~\tau $, (b) $ t \simeq 104~\tau $, (c) $ t \simeq 130~\tau $, (d) $ t \simeq 192~\tau $.
For cases above the CAC, the time sequences are: (e) $ t \simeq 108.75~\tau $, (f) $ t \simeq 143.5~\tau $, (g) $ t \simeq 161.0~\tau $, (h) $ t \simeq 175.25~\tau $.
\pd{In} our \pd{simulations} 
hole emergence for pure water occurs at approximately $ t \simeq 91~\tau $. For surfactant-laden droplets, holes (cavities) emerged at approximately $ t \simeq 100~\tau $ below the CAC \pd{(not shown)} and $ t \simeq 109~\tau $ above the CAC.} 
}
\end{figure}

\vio{In all cases, we observed the appearance of vacuum holes (cavities) during the runs at velocities around $2.2415 ~\sigma / \tau$. In Figure~\ref{fig:fig7}, the disk is depicted for pure water (Fig.~\ref{fig:fig7}a-d) and surfactant-laden droplets above the CAC (Fig.~\ref{fig:fig7}e-h), at a velocity of $V=2.2415 ~\sigma / \tau$.
The plot illustrates the thickness of the disk and the geometry of the holes. It demonstrates that, in the case of pure water, there is a stronger flow towards the rim of the disk, leading to a thinner middle region and larger holes. Conversely, in the case of surfactant-laden droplets, there is weaker flow, resulting in a thicker middle region and smaller holes.
The high kinetic energy of the expanding disk leads to the thinning of the middle region of the disk and the formation of holes. Moreover, in water droplets and droplets below the CAC, lower viscosity results in less damping of kinetic energy, which contributes to the appearance of larger holes. Finally, for the simulations conducted at a velocity of $2.4662 ~\sigma / \tau$, the viscous dissipation energy is insufficient to stabilize the disk, causing it to break apart or fragment.}
\vioo{

Moreover, to provide further insights corresponding to lower-velocity conditions, Figure S1 and Figure S2 in the ESI present figures similar to Figure~\ref{fig:fig6}, which show the disk thickness profiles for collisions at a velocity of $V = 1.5690~\sigma/\tau$ for pure water and surfactant-laden droplets (above the CAC), respectively, at comparable time instances. These lower-velocity cases correspond to conditions where no hole emergence is observed. It is evident that, for pure water, the collision energy dissipates more rapidly, leading to a thinner and less stable disk. In contrast, the presence of surfactant results in a more uniform and stable disk structure. However, at higher velocities, the emergence of holes leads to a less uniform disk, even in the case of surfactant-laden droplets (Figure~\ref{fig:fig6}e–g).
To illustrate this clearer, we have provided a movie as supplementary material (see ESI), which shows the head-on collision of water droplets (top panel) and surfactant-laden droplets above the CAC (bottom panel) at a velocity of $V = 2.2415~\sigma/\tau$. Two different views are presented: a side and a disk view (along the $x$-axis) to illustrate the evolution of the disk. The movie clearly highlights the significant differences in disk dynamics between pure water and surfactant-laden droplets. The most notable difference is that, in the case of pure water, following the collision, the droplets deform into flattened disk that expand to it's maximum diameter. Thereafter, the disk contracts, and the system undergoes damped oscillations ('beating') until the kinetic energy is dissipated. However, such oscillations and beating are absent in the case of surfactant-laden droplets, which is attributed to enhanced energy dissipation.}

\subsection{Off-Centre Collision}

Moving further, in this section we analyse off-centre collisions
where two droplets collide with an offset distance 
($h$ in Figure~\ref{fig:1}f) between their centres of mass.
These collisions are similar to head-on collisions, 
except that they involve rotational motion of the merged mass.
According to our simulations, three different scenarios may occur: 
coalescence, separation, and splattering.
Separation leads to stretching (Figure~\ref{fig:1}g) 
and can result in the formation of satellite droplets 
(Figure~\ref{fig:1}h).
Splattering occurs when the velocity 
is high and the viscous dissipation energy
cannot fully absorb the kinetic energy. 
As a result, the system falls apart, 
producing many satellite droplets (Figure~\ref{fig:1}i).

\begin{figure}[bt!]
\includegraphics[width=1.0\columnwidth]{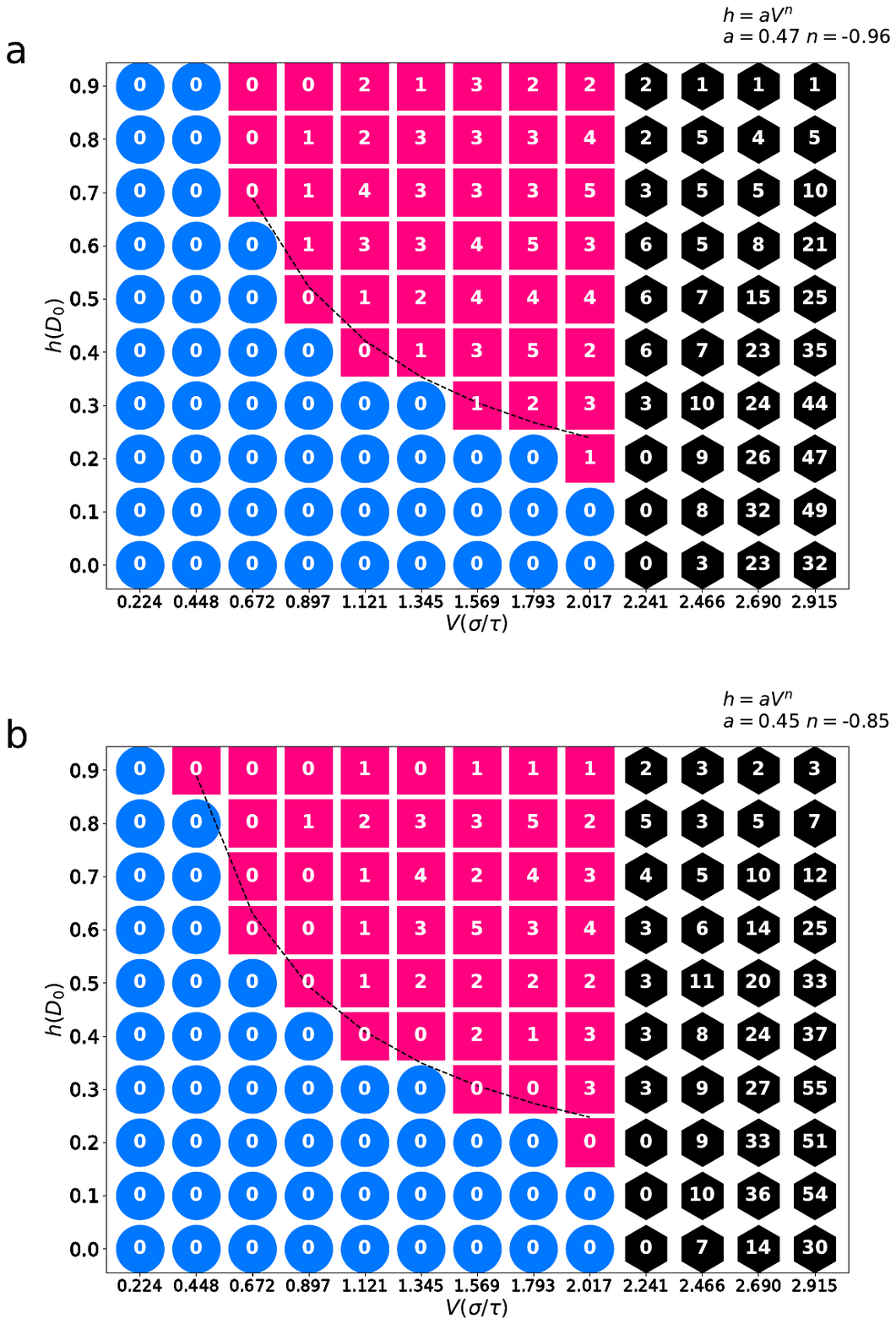}
\caption{\label{fig:fig8} State diagrams illustrating the outcomes of (a) water droplet and (b) surfactant-laden (above CAC) droplet collisions. The diagrams show the regions corresponding to different collision outcomes: coalescence (blue circles), separation (red squares), and splattering 
(black hexagons). The data have been mapped as a function of collision parameters: velocity and offset. The boundary between coalescence and separation, estimated with a power law fit with parameters given in the top right corner, 
is indicated by a dotted black line. 
 Each symbol on the diagram is labelled with a number representing the number of satellite droplets produced. 
}
\end{figure}

Figures~\ref{fig:fig8}a and b present collision outcomes for 
water and surfactant-laden droplets, respectively,
in the form of state diagrams. Higher 
velocities and larger offsets lead to a higher 
probability of separation. In each case, the number 
inside the symbol indicates the number of 
satellite droplets produced during separation. 
The boundary between coalescence and separation is 
estimated as a black dashed line.
At low velocities, coalescence is the dominant outcome. 
For low velocities $V\lesssim 0.8~\sigma / \tau$ and large offsets $h/D_0\gtrsim 0.5$, 
careful comparison of the two phase diagrams shows that the water droplets have a greater tendency to coalesce, due to the higher surface tension in 
comparison with surfactant-laden droplets.
Moreover, the tendency for 
coalescence weakens at higher velocities.
When the offset is large and the droplets just touch each 
other's surfaces, surface tension becomes more important
and can determine the fate of the collision. 
However, at lower offsets, viscosity plays a 
larger role in determining the outcome as it can dissipate more kinetic energy. 

Regarding the number of satellite droplets, 
water collisions generally produce more satellite droplets,
which is presumably again due to the lower viscosity.
In both cases, in runs with velocities above $V=2.2415  ~\sigma / \tau$, it is found that
splattering occurs, and generally more 
fragments are produced again in the case of water droplets.
 To be more precise, for the case of
$V = 2.2415~\sigma / \tau$, with a small offset (h < 0.3), the process can still
be considered coalescence. Although several very tiny satellite
droplets are formed, they are not counted as satellites due to
their extremely small size, and the two droplets merge, as
shown in the movie in the ESI.

\section{Conclusions}
\label{conclusions}

In this study, we have analyzed various properties 
during the collision of water droplets and surfactant-laden droplets, 
revealing the collision dynamics through energy balance, bridge growth rate, 
and satellite droplet counts to quantify the high energy collisions and splattering. We observe that 
in the case of coalescence, the bridge grows according 
to a power-law regime, while in the case of collision, 
an initial linear regime emerges, followed by a 
power-law regime. 
This linear regime grows and becomes more dominant by increasing the collision velocity,  and above 
a certain velocity ($V\simeq 0.23~\sigma / \tau$ for water 
and $V \simeq 0.45~\sigma / \tau$ for surfactant-laden droplets in our study) 
the power-law growth is not present
and the bridge growth is linear from the outset.
We further analysed the later time disk dynamics, 
which appears to be dominated in surfactant-laden droplets by the
effect of viscosity. 
The lower viscous dissipation energy in the case 
of water correlates with the creation of a larger disk and 
the appearance of larger holes. 
The lower energy dissipation in the case of water droplets
is also to be held responsible for the oscillation of the 
disk dimensions in time, a phenomenon not observed in 
surfactant-laden droplets. 
As a major component of the study, we quantified
offset collisions for both cases, 
and a detailed phase diagram was created to compare 
the different possible outcomes: coalescence, separation, 
and splattering, and in particular to locate the parameters
at which the behaviour undergoes a change. 
At small offsets, coalescence transitions immediately to 
splattering above a threshold velocity.
Moreover the onset of splattering appears to be independent
of offset and happens even with head-on collisions.
At the highest offsets $h\gtrsim 0.7D_0$,  the number of 
satellite droplets arising from splattering drops suddenly 
(presumably the majority of the droplet fluid passes by without significant interaction). 
Satellite droplets are also much rarer in the separation 
regime when offset is large and velocity fairly slow.
In addition, it was observed that the number of 
satellite droplets is higher in the case of water 
droplet collisions. 
\panos{Finally, we have discussed the regimes of cavity formations
in the disk-like structures and elucidated their characteristics, such 
as their thickness.}
The findings reported here provide overall a broadened
understanding of the conditions that 
lead to different collision outcomes for both water 
and surfactant-laden droplets.

\FloatBarrier

\section*{Conflicts of interest}
There are no conflicts to declare

\section*{Acknowledgements}
This research has been supported by the 
National Science Centre, Poland, under
grant No.\ 2019/34/E/ST3/00232. 
We gratefully acknowledge Polish high-performance computing infrastructure PLGrid (HPC Centers: ACK Cyfronet AGH) for providing computer facilities and support within computational grant no. PLG/2024/017543.



\balance

\renewcommand\refname{References}

\bibliography{rsc} 
\bibliographystyle{rsc} 

\end{document}